# Improving the operational stability of thermoelectric Zn$_4$Sb$_3$ by segmentation

*Peter Skjøtt Thorup[1], Christian M. Zeuthen[1], Kasper Borup[1] and Bo Brummerstedt Iversen[1*]*

[1]Center for integrated Materials Research, Department of Chemistry and iNANO, Aarhus University, DK-8000 Aarhus, Denmark

E-mail: bo@chem.au.dk

The mixed ionic-electronic conductor *β*-Zn$_4$Sb$_3$ is a cheap and high performing thermoelectric material, but under operating conditions with a temperature gradient and a running current, the material decomposes as Zn readily migrates in the structure. Here, we report an improved stability of *β*-Zn$_4$Sb$_3$ by introducing ion-blocking interfaces of stainless steel to segment the sample, produced by a rapid one-step Spark Plasma Sintering synthesis. The stability of the samples is tested under temperature gradients and electric currents, which reveals significantly improved stability of the segmented samples compared to unsegmented samples. The segmented samples are stable under temperature gradient from 250°C to room temperature with no external current, whereas the unsegmented sample decomposes into ZnSb and Zn under the same conditions. The thermoelectric figure of merit, *zT,* of the segmented sample is slightly reduced, mainly due to the increased thermal conductivity. In conclusion, a rapid one-step synthesis of segmented *β*-Zn$_4$Sb$_3$ is developed, which successfully improves the long-term operational stability by blocking the Zn ion migration.

## 1. Introduction

Thermoelectric (TE) materials are able to directly convert heat into electricity and one of the proposed components of the shift into sustainability, as TE modules are able to utilize low-value waste heat for generating high-value electricity. In low power applications, TE modules can be used in sensor technologies relevant for the Internet of Things,[1] or possibly in self-powered biomedical devices.[2] TE power generators have the huge benefit of no moving parts, no emission and thereby little to no maintenance is needed. The viability of TE power generators scales with the thermoelectric figure of merit, $zT = (\alpha^2 \sigma/\kappa)T$,[3] with $\alpha$ being the Seebeck coefficient, $\sigma$ being the electrical conductivity, $T$ the absolute temperature, and $\kappa$ the thermal conductivity, sometimes separated into the sum of the lattice ($\kappa_L$) and electronic ($\kappa_e$) thermal conductivity. As the electronic thermal conductivity is related to the electrical conductivity through Wiedemann-Franz law,[4] one of the simplest ways of improving $zT$ is lowering $\kappa_L$ without deteriorating the electrical properties. This approach gives rise to the phonon-glass electron-crystal (PGEC) concept,[5,6] where one aims to achieve glass-like phonon conduction while facilitating electron conduction like in a crystal.

Materials of mixed ionic electronic conductors (MIECs) have recently gained attention in the TE field as the ionic conduction promotes disorder in the structure that lowers the thermal conductivity.[7,8] Low thermal conductivity is observed in many superionic conducting MIECs based on Zn, Cu or Ag, such as $Cu_2X$ (X = S, Se, Te),[7] $Zn_4Sb_3$,[9] $Ag_2Te$,[10] $AgCrSe_2$,[11] and Cu- and Ag-based argyrodites,[12,13] leading to very high thermoelectric $zT$s reported for several of these superionic compounds.[8] Despite promising TE performance the main hindrance for real applications lies in the lacking stability of MIECs under working conditions of a temperature gradient and a running current. The migration of ions in the materials leads to changes in the materials composition and possibly decomposition, which significantly degrades the TE performance. Addressing these stability issues in TE devices based on MIECs is essential before industrial applications are feasible.[14]

The material $Zn_4Sb_3$ is a well-known MIEC that has received immense attention due to its excellent $zT$ above 1 in the intermediate temperature range of 130-400°C,[15] while consisting of cheap and non-toxic materials. The β-$Zn_4Sb_3$ phase crystallizes in the $R\bar{3}c$ space group with a complex structure composed of an ordered, rigid Sb lattice and a defect main lattice of Zn as well as a disordered sublattice of Zn at interstitial sites. The main lattice of Zn has a considerable number of vacancies leading to a composition of $Zn_{3.83}Sb_3$.[9] The complex sublattice of Zn ions is believed to significantly contribute to the extremely low thermal conductivity.[9,16–18] According to the phase diagram, the $β$-$Zn_4Sb_3$ phase is stable up to 477°C, where it has a phase transformation into the high temperature $γ$-phase.[19] However, it is widely known that the β-$Zn_4Sb_3$ decomposes already well below 492°C,[15,20,21] and it is proposed that the onset of Zn migration is at temperatures above 152 °C.[22] The main decomposition reaction mechanism for $Zn_4Sb_3$ is $Zn_4Sb_3 \rightarrow$ 3 ZnSb + Zn at temperatures from 200°C, which is observed regardless of atmosphere, synthesis method and doping content.[23–25] This decomposition is further accelerated by a thermal gradient or electric current, where the highly mobile $Zn^{2+}$ ions in $Zn_4Sb_3$ migrates down the resulting electric field gradient leading to decomposition into ZnSb in the Zn poor region (positive electric potential) and growth of Zn whiskers in the Zn rich region (negative electric potential).[26–29] The irreversible decomposition leads to the deterioration of the TE properties and the growth of Zn whiskers ruins the electrical contacts, and alters the geometric shape of the TE leg, deteriorating the whole TE module. One study finds that the ion migration and decomposition of $Zn_4Sb_3$ can happen even at room temperature (RT), only under the influence of an electric current, although this was without taking Joule heating into account and without explicitly measuring the temperature of the sample.[30] The mobility of Zn ions in $Zn_4Sb_3$ is exceptionally high, with an activation energy of only 11 kJ mol$^{-1}$.[31] This is lower than Cu$^+$ in $Cu_2Se$ at 13.5 kJ mol$^{-1}$,[32] and even approaches the value of 9.6 kJ mol$^{-1}$ for the fast ion conductor of α-AgI.[33]

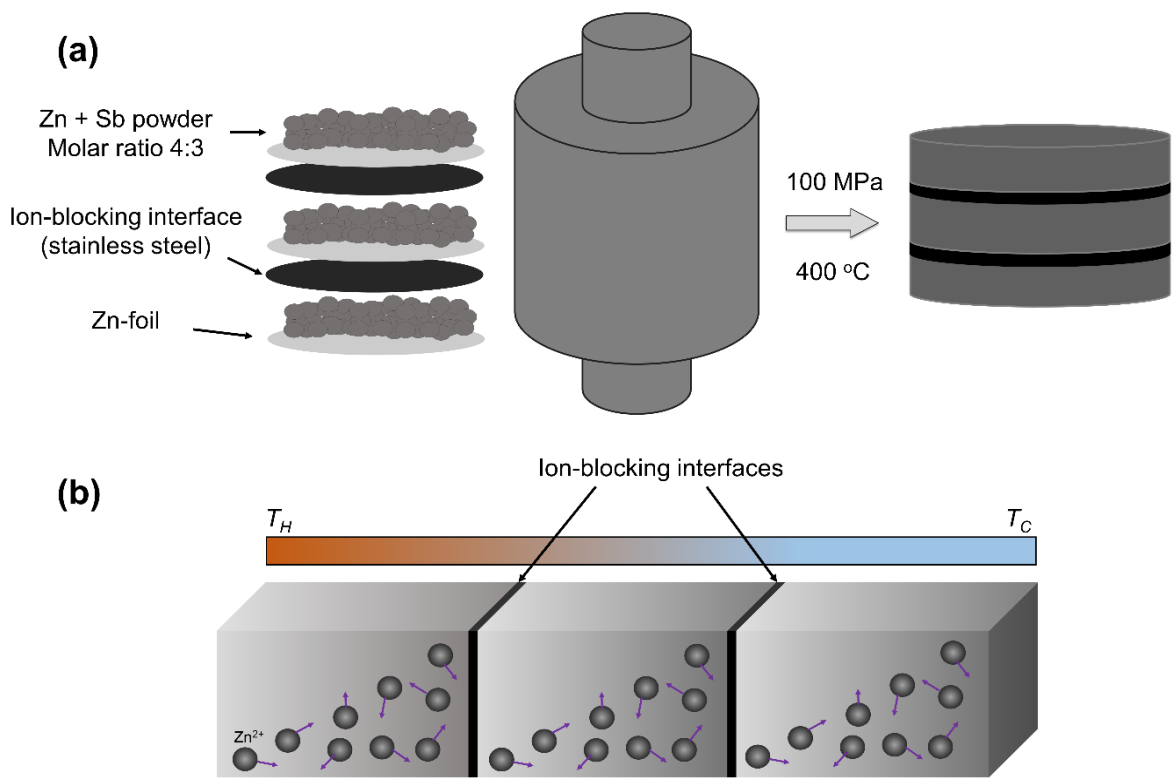

**Figure 1.** a) One-step synthesis procedure to fabricate segmented $Zn_4Sb_3$ with Spark Plasma Sintering using stainless steel as the ion-blocking interface. b) Under working conditions of a temperature gradient and applied field, a zinc ion concentration gradient will form inside each segment. The ion concentration will be reset at each ion-blocking interface, ideally being below the upper limit for Zn metal formation at all times.

Attempts to suppress the Zn migration to improve the stability of $Zn_4Sb_3$ have been manifold, and include doping with e.g. Mg,[34,35] Cd,[23] and Ag,[36] nanostructuring,[35,37,38] inclusions of ZnO[39] and

TiO$_2$ nanoparticles,[39,40] and formation of amorphous grain boundaries,[41] but unfortunately all attempts so far have led to only insufficient improvements.

The MIECs of copper chalcogenides, and especially Cu$_2$Se, have been extensively researched for use in TE devices, but the nature of the highly mobile Cu$^+$ ions leads to Cu metal formation, Se evaporation, changes in contact resistances over time, and structural degradation. The material stability is a problem that is still not solved despite numerous attempts.[8,42–46] Recently, it has been shown that segmentation of Cu$_{2-\delta}$S samples with ion-blocking interfaces can increase the stability compared with unsegmented samples, by withstanding higher current densities under isotherms and thermal gradients.[47] By segmentation, the total critical voltage needed for Cu metal formation is increased, as the voltage across each segment is reduced.[47,48] As the ion conduction of Zn$^{2+}$ ions resembles that of Cu$^+$ ions in Cu$_2$S and Cu$_2$Se, the segmentation approach may also improve the stability of Zn$_4$Sb$_3$.

In this article, it is shown that the stability of Zn$_4$Sb$_3$ can be substantially improved using segmentation with ion-blocking interfaces. A fast, direct one-step synthesis of segmented Zn$_4$Sb$_3$ using Spark Plasma Sintering (SPS) is developed, with stainless steel as the ion-blocking interface (Figure 1a). These interfaces reset the chemical potential, ideally keeping the material stable under operating conditions, provided that the chemical gradient inside each segment does not reach an upper limit where metal segregation starts to occur (Figure 1b). The stability of both segmented and unsegmented Zn$_4$Sb$_3$ samples are tested using the Aarhus Thermoelectric Operando Setup (ATOS),[29] which is able to apply a current and temperature gradient, resembling working conditions in a thermoelectric module. The samples are examined with Potential Seebeck Microprobe (PSM) scans after the stability tests, to gain insight into the Zn migration in the structure and decomposition of the Zn$_4$Sb$_3$ phase into ZnSb and Zn, as the Seebeck coefficient is very sensitive to different phases.

We demonstrate improved stability of the segmented Zn$_4$Sb$_3$ as compared to pure samples, under conditions of an isotherm or temperature gradient, both with or without applied external current. Thermoelectric property measurements reveal a higher power factor and a higher thermal conductivity of the segmented samples compared with the pure samples, resulting in only a slight reduction in the figure of merit $zT$ at temperatures up to 300°C.

## 2. Experimental Section/Methods

*Synthesis*: The synthesis of segmented Zn$_4$Sb$_3$ is built upon the fast, direct synthesis and compaction using SPS pressing, developed previously by Yin *et al*[49,50]. Zinc powder (99.99%, grain size <45 μm,

MERCK KGaA) and antimony powder (99.5%, grain size <150 μm, SIGMA-Aldrich CHEMIE GmbH) were weighed in a 4:3 molar ratio. The powders were mixed for 2x15 min in a ball mill mixer (SpectroMill, CHEMPLEX INDUSTRIES, INC). Three layers of 6.5 g powder, separated by 0.05 mm stainless steel (grade 304) foil, were loaded into a graphite die with a diameter of 2.56 cm, resulting in three segments with a height of 2.2 mm each. To compensate for Zn migration during synthesis a 0.1 thick mm Zn foil (0.44 g) was placed at the bottom of each segment, and the die was protected by BN spray. The pellet was sintered for 5 min at 400°C and 80 MPa using an SPS-515 instrument (SPS SYNTEX INC, Japan).

*Material characterization:* The purity of the sample was checked by PXRD measurement on the top and bottom of a gently polished pellet using a Rigaku Smartlab diffractometer equipped with a Cu K$\alpha_1$ source and Bragg-Brentano optics. To identify the phases, Seebeck microprobe scanning was performed on cross-sectional areas of synthesized pellets, and the surface of bars after stability tests, using a commercial PANCO PSM[51]. The contact resistance over the ion-blocking interfaces was measured using the potential measurements with the PSM, on a bar-shaped sample with cross-sectional area of 0.74 mm$^2$.

*Physical property measurements:* The electrical resistivity and Seebeck coefficient is measured between 50°C and 300°C in steps of 50°C with an Ulvac ZEM-3. The volume fraction of stainless steel in the measured region between the probes is 1.7%, which is 0.7% higher than the volume fraction of stainless steel of 1.0% in the entire segmented leg.

The thermal diffusivity $\alpha$ is measured between 50°C and 300°C in steps of 50°C using a Netzsch LFA 467 HT, on a segmented pellet with only one stainless steel interface. The volume fraction of stainless steel in the measured sample is 2.4%. The heat capacity ($C_p$) was determined using a reference sample of Pyroceram 9606, and the thermal conductivity then calculated according to the equation $\kappa = \alpha d C_p$.

*Stability test setup:* For stability tests, the home-built ATOS (Aarhus Thermoelectric Operando Setup) is used, previously described in detail.[29] For the tests, the temperature was ramped to the set value within 5 minutes, and subsequently the current was turned on after an additional 5 minutes, which allowed the temperature to stabilize.

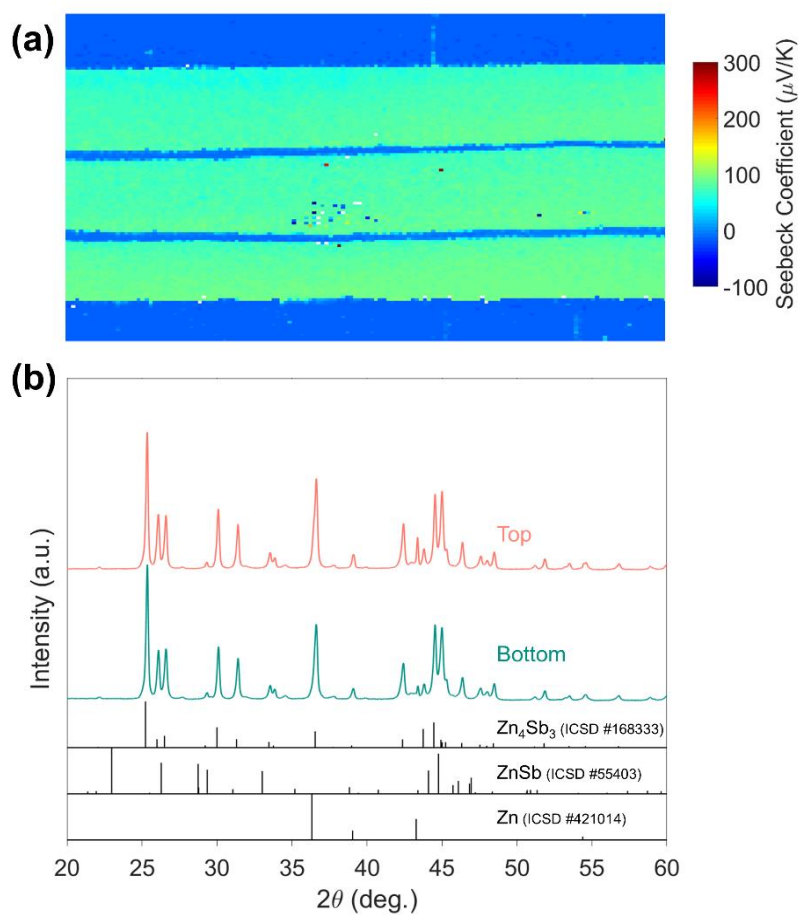

**Figure 2.** (a) Cross-sectional PSM scan of SPS pressed segmented pellet of $Zn_4Sb_3$ with a Zn foil at the bottom of each segment. (b) PXRD data at the top and bottom of a three-segment pellet synthesized with a Zn foil at the bottom of each segment.

## 3. Results and discussion
### 3.1. Phase characterization of synthesized segmented pellet

Previous studies of segmented MIEC's have used time-consuming preparation methods, which are challenging to scale for industrial fabrication.[47] For segmentation to be a viable method of stabilizing TE MIECs, the fabrication must be simplified. For $Zn_4Sb_3$, Hao *et al.*[50] have shown a rapid one-step bulk synthesis method using SPS. The challenge of using SPS is that the $Zn^{2+}$ ions migrate along the current direction, thereby depleting one side of the pellet of Zn, resulting in decomposition into ZnSb. They further showed that compensating for the migration by introducing a Zn foil at the bottom of the pellet resulted in phase pure $Zn_4Sb_3$ samples. This is a very rapid and scalable synthesis method, and it is well suited for synthesizing segmented samples.

First, a good ion-blocking interfacial layer is identified. In this study, stainless steel was chosen due to availability, good mechanical stability, price and low solubility and reactivity with Zn. To test the ion-blocking capabilities, a sample was synthesized using only the Zn-Sb precursor mixture and ion-blocking interface. As no additional Zn was introduced it was expected that the sample should partially decompose into ZnSb due to migration of Zn. If the ion-blocking interface was effective, each segment should have ZnSb in the bottom and $Zn_4Sb_3$ in the top. The results are shown in the supporting information Figures S1a and S2. They clearly show that stainless steel is indeed a sufficient ion-blocking interface, as each segment contains ZnSb at the bottom and $Zn_4Sb_3$ at the top.

To achieve phase pure samples, an additional Zn source needs to be present in the SPS synthesis. Unfortunately, it is not enough introducing Zn foil at the bottom on the entire pellet, as the stainless steel ion-blocking interface naturally also blocks the Zn from the foil from compensating the following segments. Therefore, an additional Zn source needs to be present in each segment, which is done by introducing Zn foil at the bottom of each segment. This results in a phase pure sample of $Zn_4Sb_3$ with no ZnSb. The homogeneity was examined by Seebeck coefficient maps (Figure 2a), revealing a sample consisting solely of the $Zn_4Sb_3$ phase (80-100 µV/K).[50] As Zn migrates during SPS synthesis, a small gradient in Zn concentration is present in the sample, as evident from PSM with a tighter color scale (Figure S1b). PXRD shows that small amounts of excess Zn exists in the sample (Figure S2), which might be eliminated by closely controlling the stoichiometry. Careful examination of this should be the focus of an additional study.

**3.2 Stability tests under working conditions**

The stability of bars of $Zn_4Sb_3$ is tested under realistic working conditions of a temperature gradient, both with and without applying an external current. The stability of the segmented samples is directly compared with the stability of unsegmented samples, referred to as *pure* from here on, under the same operating conditions. The geometrical size of the samples is kept constant in all experiments to minimize the parameter space. The length of the bars is kept at 6.2 mm, with the electrodes and Ni probes separated by 4.2 mm, and the cross-sectional area is kept at 0.80 mm². The decomposition of $Zn_4Sb_3$ into ZnSb and Zn is monitored directly by a rise in resistance across the sample, as the ZnSb phase has a substantially higher resistivity than the $Zn_4Sb_3$ phase.[15] The resistance across the sample is measured using four-point geometry, thus the position of the Ni probes is important, as a change in the measured resistance is only detected when the ZnSb phase propagates between the probe tips, illustrated by the schematic of the setup (Figure S3).

**3.2.1 Stability tests with temperature gradient and without external current**

The stability of segmented versus pure $Zn_4Sb_3$ was evaluated using three different temperature gradients $\Delta T = 350°C – RT$, $\Delta T = 300°C – RT$ and $\Delta T = 250°C – RT$. From the measured resistance of segmented and pure $Zn_4Sb_3$ samples under these various temperature gradients the decomposition event into ZnSb and Zn is distinctly visible with the sharp increase in resistance (Figure 3a). In the temperature gradient of $\Delta T = 350°C – RT$ the time before decomposition for the segmented sample is roughly increased twofold compared with the pure sample, with Zn whiskers visibly developing at each ion-blocking interface (Figure S4). In the temperature gradient of $\Delta T = 300°C – RT$ no clear improvement in the time until decomposition is observed. However, a clear improvement in stability is observed with $\Delta T = 250°C – RT$, with the pure sample decomposing after 30 h while the segmented sample does not exhibit signs of decomposition for one week (165 h) of testing. A PSM scan, performed after the stability test, reveals a stable segmented sample, with no signs of decomposition into the ZnSb phase (Figure 3b), while the pure sample is clearly decomposed close to the hot end of the sample (Figure 3c). This is a strong indication of a steady state regime reached in the segmented sample in a temperature gradient of $\Delta T = 250°C – RT$, where the opposing electric field from the Zn ion concentration cancels the directional force generated from the temperature gradient and ensures zero net flux of $Zn^{2+}$ ions, which stabilizes the $Zn_4Sb_3$ phase. This is not accomplished in the pure $Zn_4Sb_3$ sample subjected to the same conditions.

**3.2.2 Stability tests with external current**

The stability of segmented versus pure $Zn_4Sb_3$ was evaluated with an applied external current density of $J = 0.5$ A/mm$^2$, which resembles a realistic current density in a TE module, under temperature gradients of $\Delta T = 250°C – RT$, $\Delta T = 200°C – RT$, and an isotherm $\Delta T = 250°C – 250°C$. All the experiments are performed in power generation mode, *i.e.* the applied current is in the direction dictated by the Seebeck coefficient.[29] From the measured resistance across the sample (Figure 4a) the time before decomposition is increased from 3 h in the pure sample to 14 h in the segmented sample under $\Delta T = 250°C - RT$. When decreasing the temperature gradient to $\Delta T = 200°C - RT$ the pure sample decomposes after 9 h while the segmented sample exhibits no increase in resistance after one week (165 h) of testing. The extend of decomposition is examined with PSM scans and shown for the samples with $\Delta T = 200°C - $ RT (Figure 4b,c) and $\Delta T = 250°C - $ RT (Figure S5). The pure sample has decomposed in the hot end of the bar, where the current enters. Despite no rise in the resistance, the segmented sample has undeniably started to break down in the first segment, which is explained by the ZnSb phase front not reaching the probe tip within the timeframe of the test. However, there is no sign of decomposition present in the second and third segment. In the right side of the second segment the Seebeck coefficient is slightly lower, which indicates a Zn rich region in the sample. This shows that the excess Zn in the sample is still able to migrate in this segment, but the conditions are not sufficient for decomposition of $Zn_4Sb_3$ into ZnSb. This is explained by the considerable phase width of $Zn_4Sb_3$.[52,53] The main factor is most likely the temperature being too low in this segment, while a minor factor is the increased Zn concentration in the cold end impeding further movement of Zn and stabilizing the $Zn_4Sb_3$ phase. With both electrode heaters set to 250°C and an external current density of $J = 0.5$ A/mm$^2$, the time before decomposition in the segmented sample is increased from 1 h to 22 h (Figure 2a). This indicates that the segmented samples are not merely more stable due to a reduced temperature gradient inside each segment, but that the potential is indeed lowered within each segment, giving rise to improved stability. It should be noted, that due to convective cooling across the bar and low thermal conductivity in the sample, the temperature will be lower in the center of the sample than at the end of the bars in direct contact with the electrodes. The thermal contact resistance of the ion-blocking interfaces possibly enhances this effect. From PSM scans (Figure S6) it is observed that the segmented sample is decomposed in every segment, indicating the importance of a certain temperature needed for the decomposition into ZnSb.

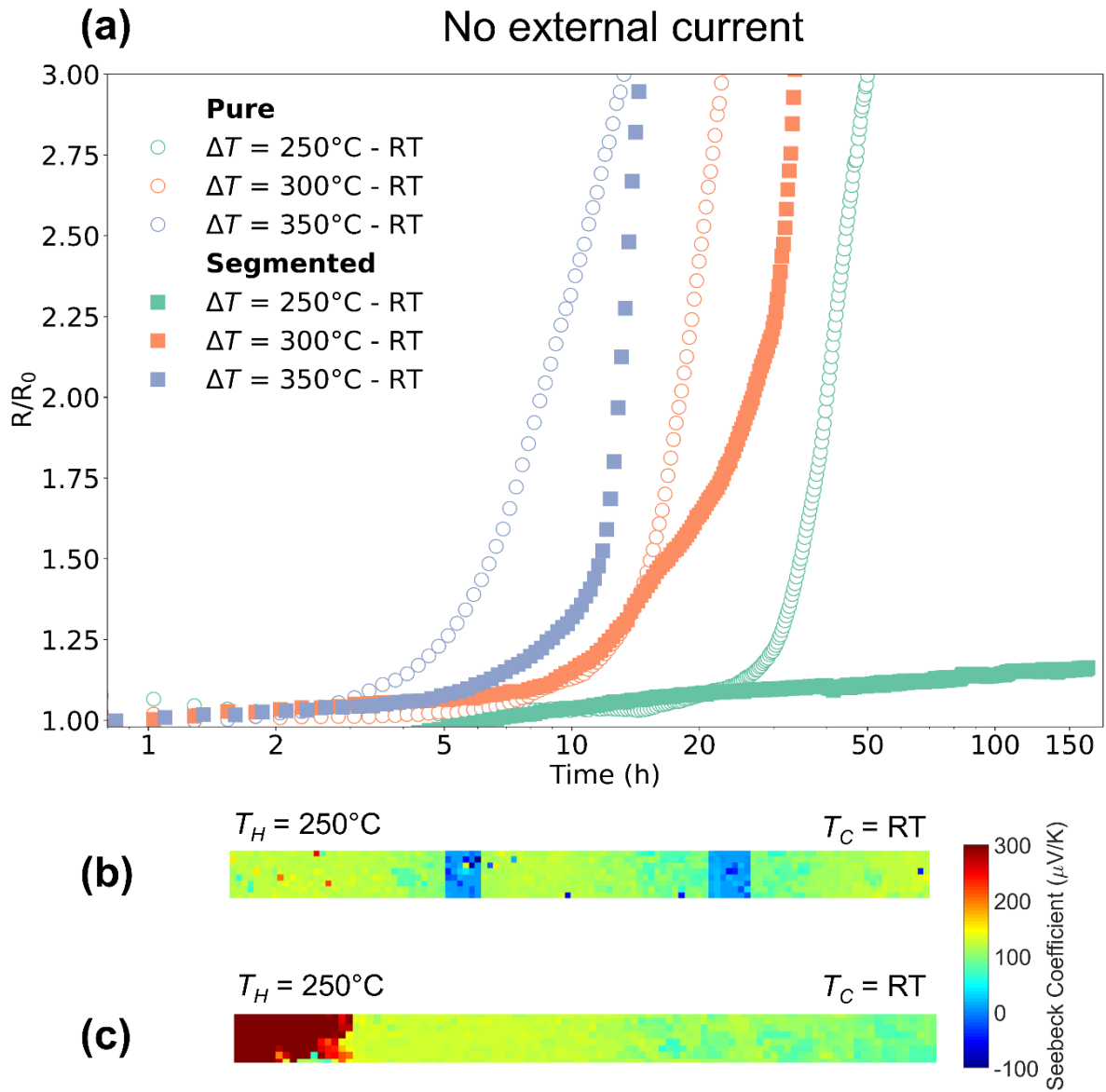

**Figure 3.** a) Normalized resistance as a function of time for segmented and pure $Zn_4Sb_3$ samples under various temperature gradients without external current, plotted with a logarithmic time axis. A sharp rise in resistance indicates decomposition into the ZnSb phase. PSM scan for the segmented (b) and pure sample (c) after test at $\Delta T = 250°C - RT$. No decomposition is observed for the segmented sample, while the pure sample has decomposed into ZnSb (red region).

The stability tests establish that the segmented $Zn_4Sb_3$ samples are stable for significantly longer times than the pure $Zn_4Sb_3$ under all tested conditions of temperature gradients, both with or without an external current.

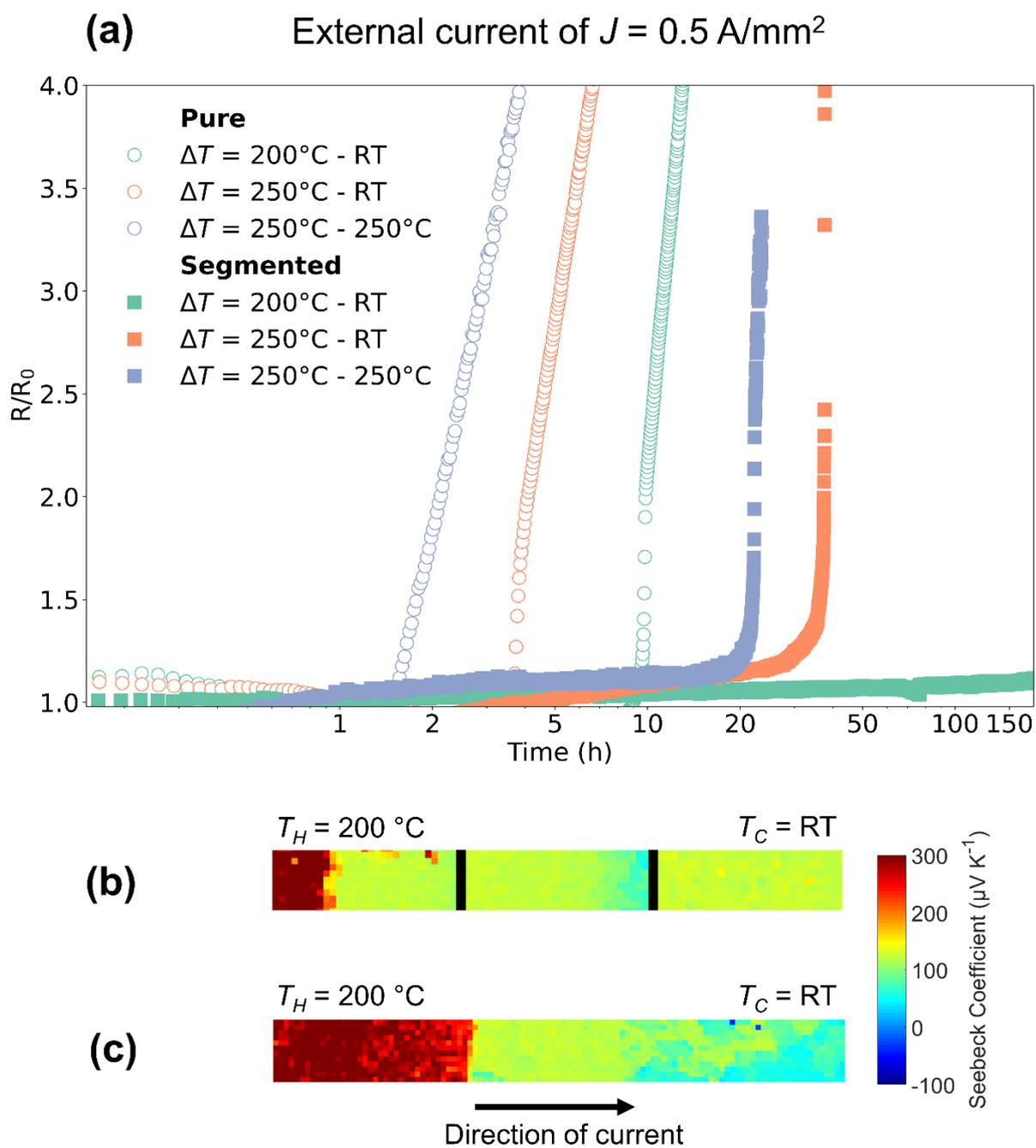

**Figure 4.** a) Normalized resistance as a function of time for segmented and pure $Zn_4Sb_3$ samples under various temperature gradients with an external current of $J = 0.5$ A/mm$^2$. A sharp rise in resistance indicates decomposition into the ZnSb phase. Post PSM scan for the segmented sample (b) and pure sample (c) with $\Delta T = 200\ ^{\circ}\text{C} - \text{RT}$ and $J = 0.5$ A/mm$^2$. The segmented sample is measured in three different instances, with the black bars indicating the position of ion-blocking interfaces.

## 3.3 Thermoelectric properties

The thermoelectric properties of the synthesized segmented $Zn_4Sb_3$ sample are compared to the properties of pure $Zn_4Sb_3$, to ensure that the approach of introducing metallic ion-blocking interfaces does not have a detrimental impact on the performance. All measurements are performed perpendicular to the stainless steel interfaces. The measured resistivity (Figure 5a) reveals a higher conductivity at all temperatures for the segmented sample, while the Seebeck coefficient (Figure 5b) is slightly lower for the segmented sample, as expected due to the metallic stainless steel interface and minor Zn phase, that has a lower resistivity and Seebeck coefficient than $Zn_4Sb_3$. This affects the power factor $\alpha^2/\rho$ (Figure 5c), and reveals a better electrical performance of the segmented $Zn_4Sb_3$ sample compared to the pure $Zn_4Sb_3$. The thermal conductivity (Figure 5d) of the segmented sample is significantly higher than the pure sample, as expected with the high thermal conductivity of the stainless steel interfaces and Zn phase. The calculated figure of merit (Figure 5e) of the segmented $Zn_4Sb_3$ sample is slightly lower than the pure $Zn_4Sb_3$ at all temperatures with a reduction of 15% at 300°C. It should be noted that the transport property values given for the segmented $Zn_4Sb_3$ sample are not material specific values but an average of the entire segmented leg. The stainless steel interfaces do not directly contribute to the thermoelectric power generation, so the desired properties of the interfaces are both a low resistivity and a high thermal conductivity, to obtain the highest possible temperature gradient inside each segment.

The electrical contact resistance of the interfaces was determined from electric potential measurements along a segmented $Zn_4Sb_3$ using the PSM, where the resistance ($R_{pot}$) was calculated at each position as $R_{pot} = U_{pot}/I_{pot}$, with $U_{pot}$ being the electric potential and $I_{pot}$ the electric current. The average $R_{pot}$ and Seebeck coefficient across a segmented $Zn_4Sb_3$ is shown in Figure 6. The stainless steel interfaces are clearly identified by the drop in Seebeck coefficient and flattening of $R_{pot}$ (Figure 6a). From the slope of $R_{pot}$ it is evident that the contact resistance of the stainless steel interfaces is very small, and is smallest in contact with the Zn rich region of the segments. The contact resistivity ($\rho_c$) between the stainless steel interface and Zn poor region of $Zn_4Sb_3$ is calculated to $\rho_c = R_c \cdot A = 0.046$ m$\Omega \cdot$cm$^2$ (Figure 6b), where $R_c$ is the contact resistance and A is the contact area. Combining the low contact resistivity with the very low resistivity of stainless steel ($7.2 \cdot 10^{-4}$ m$\Omega \cdot$cm), the total resistivity of the segmented leg is even lower than the individual $Zn_4Sb_3$ segments.

It should be noted that these samples are composed of a Zn rich $Zn_4Sb_3$ phase which has been shown to be beneficial for both improving the power factor and lattice thermal conductivity.[25] An extra Zn phase in itself is not desired, since it increases the thermal conductivity and lowers the Seebeck coefficient.[54]

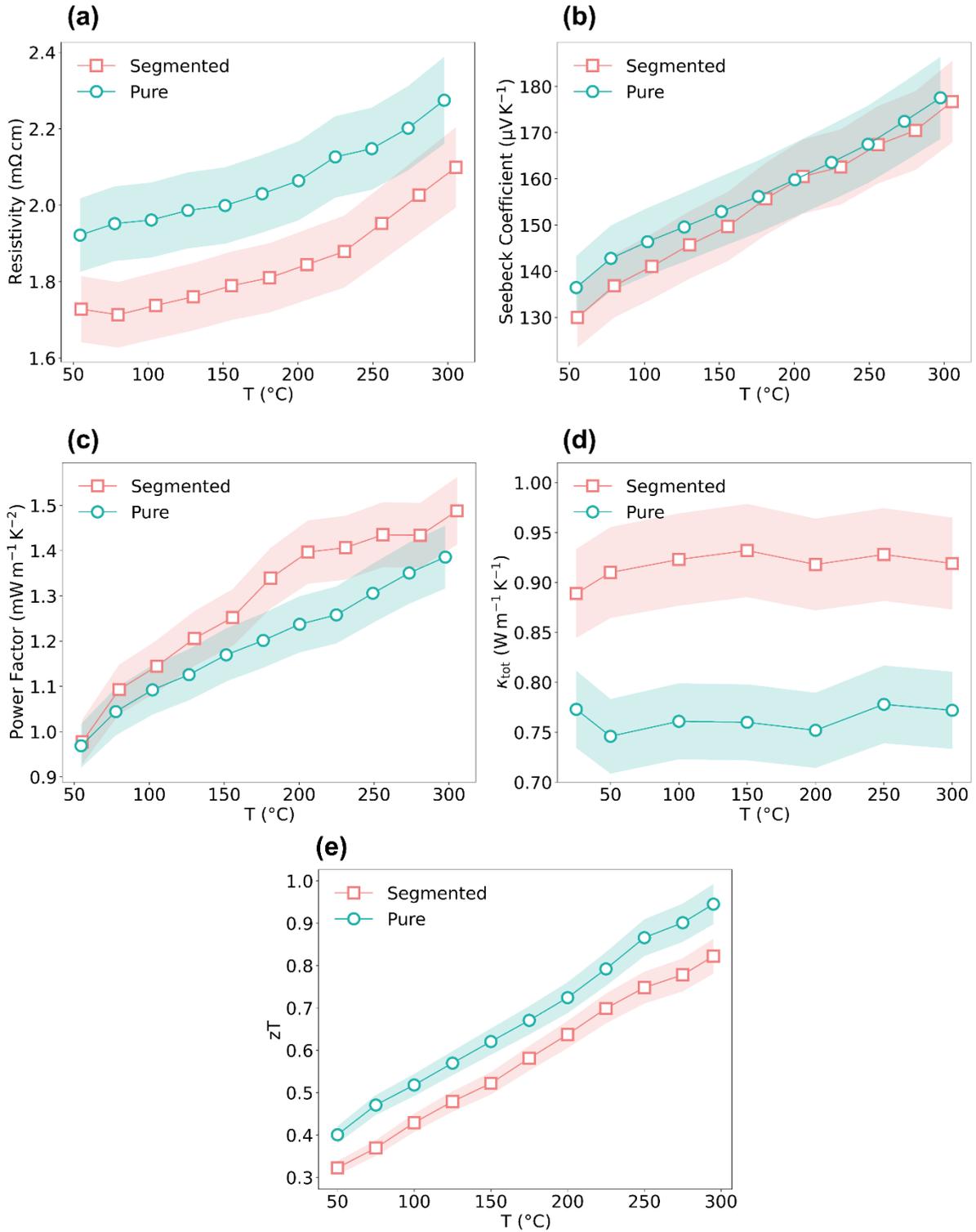

**Figure 5.** Thermoelectric properties as a function of temperature for pure and segmented $Zn_4Sb_3$ samples. (a) Resistivity, (b) Seebeck coefficient, (c) calculated power factor, (d) total thermal conductivity, and (e) calculated figure of merit $zT$. Shaded areas show 5% error margin and solid lines are linear interpolated.

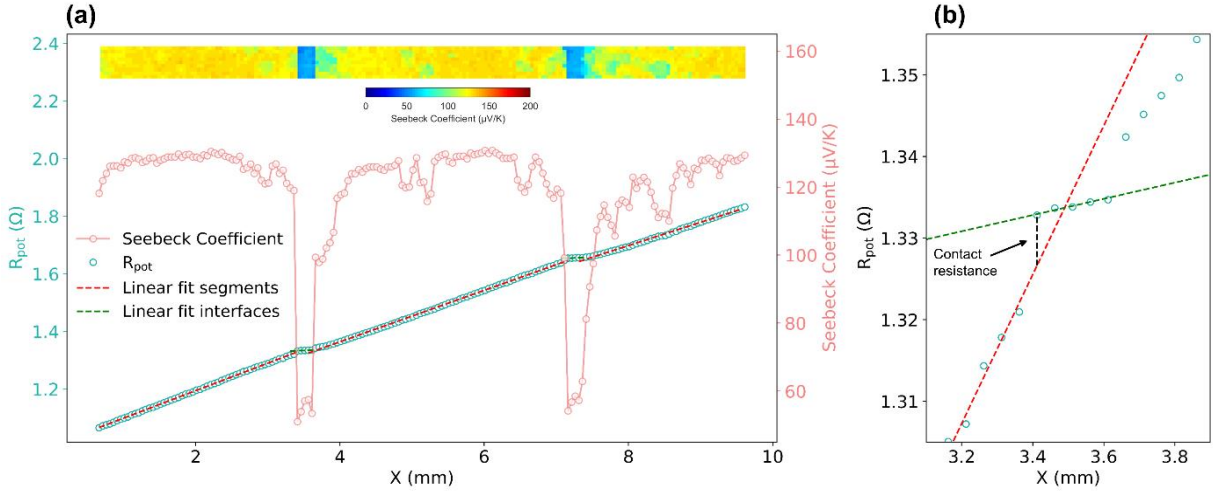

**Figure 6.** a) The average resistance, $R_{pot}$ (blue), across a segmented $Zn_4Sb_3$ sample from PSM measurements alongside the average measured Seebeck coefficient (orange). Linear expressions are fitted to the individual segments (red dotted line) and interfaces (green dotted lines). Inset is a 2D Seebeck map. b) Zoom on the first stainless steel interface where the contact resistance is visualized as the jump in resistance from the $Zn_4Sb_3$ segment to the stainless steel interface. It is evident that the contact resistance is lower in the Zn rich end of the segments (here to the right side of the interface).

The property measurements reveal a higher power factor for segmented $Zn_4Sb_3$ compared to pure $Zn_4Sb_3$, while the overall $zT$ is reduced slightly due to the increased thermal conductivity of the segmented samples. Nevertheless, segmented $Zn_4Sb_3$ still possesses an acceptable TE performance in the mid-temperature range, while obtaining improved stability.

## 4. Discussion

The segmented sample is found to be completely stable at a temperature gradient of $\Delta T = 250°C -$ RT with no external current applied, while the pure sample decomposes after 30 h. This suggests that the segmentation approach is indeed successful in improving the stability of the MIEC $Zn_4Sb_3$, and that steady state conditions can be achieved at this temperature gradient. Improvement of stability by segmentation is only significant when the sample is at, or close to, steady state conditions. The steady state regime, and thus the stability window, can be extended to higher temperature gradients and applied fields by increasing the number of segments, as this will lower the voltage across each segment further. The addition of a metallic interface material will reduce the overall $zT$, and care must be taken in choosing a suitable material of the ion-blocking interface. The fundamental role of the ion-blocking interface is to conduct electrons while blocking the movement of $Zn^{2+}$ ions, however, to avoid harming the thermoelectric properties of the material, additional requirements must be met by its physical properties. Firstly, it must have a thermal conductivity, which is higher than the thermoelectric material. If this is not met, the thermal gradient across the material will effectively be inhibited, which will lower the generated Seebeck voltage. Secondly, it must have a high electrical conductivity to facilitate the current flow and the contact resistance between the thermoelectric material and the interface material is to be minimized, to avoid losses and reduce Joule heating. Furthermore, the interface material must be mechanically stable and chemically inert under operation conditions. To obtain sufficient mechanical stability a chemical reaction between the thermoelectric material and the interface material is needed to some extent. The $Zn^{2+}$ ion migration during SPS synthesis presumably facilitate a degree of chemical reaction, that results in an excellent mechanical stability of the stainless steel interfaces in the segmented $Zn_4Sb_3$ samples. Carbon was initially used as the interface material, as suggested in the literature[47], but due to its layered structure it has poor mechanical stability and has the undesired ability to intercalate $Zn^{2+}$ ions. Stainless steel was the material of choice for the ion-blocking interface due to desired electrical and mechanical properties, and cheap cost. Other possible candidates of interface materials could be metals of Ni, Cu, Ta, Nb or Mo. Further research is needed to find the optimal material for the ion-blocking interface, and to optimize the number and thickness of the ion-blocking interfaces to achieve the best combination of stability and thermoelectric properties.

It is a critical voltage, and not a critical current density, that is the deciding factor for metallic deposition in MIECs.[47] To keep the voltage across the $Zn_4Sb_3$ phase constant, the tests in this study are performed in constant current mode, as the resistance of the $Zn_4Sb_3$ phase is approximately constant before it decomposes into ZnSb and Zn, and changes in contact resistances between the sample and the electrodes during operation will not alter the voltage gradient in the $Zn_4Sb_3$ phase. When performing stability tests with applied external current it is necessary to acknowledge the effect of Joule heating, which is most pronounced in boundaries of increased contact resistance. These are present between the sample and tower, and between the sample and the ion-blocking interfaces. It is very difficult to decouple thermal and electrical contributions to the observed stability, especially with a temperature dependent ion mobility of mixed ionic-electronic conductors. Furthermore, the true temperature along the sample is known to deviate from the temperature set by the electrode heaters in this setup.[29] The stability tests in this study performed with a temperature gradient and no external current resembles realistic operating conditions, with the voltage difference only generated by the Seebeck effect, and the effect of Joule heating being negligible when no external current applied. However, it should be noted that in a stability test of a single TE leg it is not possible to achieve both a realistic voltage difference and current density at the same time, as compared to leg in a TE module. The voltage difference generated in a test with only a temperature gradient is a worst case scenario, as the voltage will be lowered when currents flows in a TE module.

In this study, the stability of $Zn_4Sb_3$ has been evaluated from the measured electrical resistance during continuous long-term exposure to a thermal gradient and constant external current. Another approach could be to determine the critical current density and critical voltage necessary for metal segregation, by ramping the current density in steps, as reported on $Cu_{2-\delta}S$.[47] However, using this approach to determine the critical current density of $Zn_4Sb_3$ is problematic, as it can take several hours for the decomposition into ZnSb and Zn to occur. Furthermore, the repeated on and off switching of the current presumably increases the contact resistance at the interfaces. The resulting increased joule heating accelerates the decomposition, with local Joule heating clearly visible at the interfaces, while the resistance rises rapidly (Figure S7a,b). Consequently, the current density ramping test is inadequate in determining the critical current density for the segmented $Zn_4Sb_3$ samples.

## 5. Conclusion

The effect of segmentation of $Zn_4Sb_3$ on the thermoelectric properties and stability under working conditions has been studied, by introducing ion-blocking interfaces of stainless steel into the sample. A fast, direct synthesis of segmented $Zn_4Sb_3$ by SPS has been developed, and it is able to produce dense and mechanically stable samples within minutes. Stability tests under working conditions of a temperature gradient and an electric current reveals significantly improved stability for segmented samples compared to unsegmented $Zn_4Sb_3$, under temperature gradients up to 350°C to RT and current densities of 0.5 A/mm². At a temperature gradient of $\Delta T = 250°C$ – RT with no external current, the segmented sample remains stable for at least 165 h while the unsegmented sample breaks down after 30 h, which demonstrates the stabilizing effect of segmentation by obtaining steady state conditions with a low ion concentration gradient. The introduction of metallic regions in the segmented sample results in an increased power factor and a higher thermal conductivity, which overall reduces the *zT* by 15% at 300°C. The approach of using ion-blocking interfaces is shown as a proof-of-concept to be an effective tool for improving the stability of $Zn_4Sb_3$, and it opens the door for reestablishing $Zn_4Sb_3$ as an attractive thermoelectric material. The segmentation approach can be optimized even further to improve the stability window in operational conditions of temperature gradient and current densities.

**Acknowledgements**

Funding from the Villum Foundation is gratefully acknowledged.